# Extended High Circular Polarization in the Orion Massive Star Forming Region: Implications for the Origin of Homochirality in the Solar System

TSUBASA FUKUE[1*], MOTOHIDE TAMURA[1,2], RYO KANDORI[1], NOBUHIKO KUSAKABE[1], JAMES H. HOUGH[3], JEREMY BAILEY[4], DOUGLAS C. B. WHITTET[5], PHILIP W. LUCAS[3], YASUSHI NAKAJIMA[6,1] and JUN HASHIMOTO[2]

[1] *National Astronomical Observatory of Japan, 2-21-1 Osawa, Mitaka, Tokyo 181-8588, Japan.;*

[2] *Graduate University of Advanced Science, 2-21-1 Osawa, Mitaka, Tokyo 181-8588, Japan.;*

[3] *Centre for Astrophysics Research, Science and Technology Research Institute, University of Hertfordshire, Hatfield, Herts. AL10 9AB, UK.;*

[4] *School of Physics, University of New South Wales, NSW 2052, Australia.;*

[5] *New York Center for Astrobiology and Department of Physics and Astronomy, Rensselaer Polytechnic Institute, Troy, NY 12180-3590, USA.;*

[6] *Department of Astrophysics, Nagoya University, Nagoya 464-8602, Japan.*

(∗ author for correspondence, e-mail: tsubasa.fukue@nao.ac.jp)

**Abstract** We present a wide-field (~6'×6') and deep near-infrared ($K_s$ band: 2.14 μm) circular polarization image in the Orion nebula, where massive stars and many low-mass stars are forming. Our results reveal that a high circular polarization region is spatially extended (~0.4 pc) around the massive star-forming region, the BN/KL nebula. However, other regions, including the linearly polarized Orion bar, show no significant circular polarization. Most of the low-mass young stars do not show detectable extended structure in either linear or circular polarization, in contrast to the BN/KL nebula. If our solar system formed in a massive star-forming region and was irradiated by net circularly polarized radiation, then enantiomeric excesses could have been induced, through asymmetric photochemistry, in the parent bodies of the meteorites and subsequently delivered to Earth. These could then have played a role in the development of biological homochirality on Earth.

***Keywords*** *circular polarization, Orion Nebula, enantiomers, homochirality, origins of life*

# 1. Introduction

The origin of biomolecular homochirality, which refers to the phenomenon that terrestrial living material consists almost exclusively of one enantiomer, left-handed amino acids and right-handed sugars, is a longstanding mystery that is critical to understanding the origin and development of life (Bonner 1991, 1995; Meierhenrich and Thiemann 2004; Barron 2008). Amino acids in several meteorites (e.g., Murchison, Murray, Orgueil) have been found to have enantiomeric excesses (EEs) of the same handedness as that seen in biological amino acids (Cronin and Pizzarello 1997; Pizzarello and Cronin 2000; Pizzarello et al. 2003; Pizzarello et al. 2008; Glavin and Dworkin 2009; Sephton 2002). Such detection of EEs in meteorites is consistent with the hypothesis that life on Earth was seeded by the delivery of organics from outer space during the heavy bombardment phase of Earth's early history (Bailey et al. 1998; Bailey 2001; Buschermöhle et al. 2005). Furthermore, homogeneity of right-handed sugars may be also be initiated by exogenous injection of low EEs of amino acids as a catalyst (Weber 2001; Pizzarello and Weber 2004; Córdova et al. 2005; Córdova et al. 2006).

Amino acids or amino acid precursors (Botta and Bada 2002) can exist in space conditions. Amino acids were obtained in laboratory experiments that simulate ultraviolet (UV) photolysis of interstellar ice analogues (Bernstein et al. 2002; Muñoz-Caro et al. 2002; Nuevo et al. 2008). Experiments have indicated that cosmic rays can produce amino acid precursors in icy environments (Hudson et al. 2008). However, external effects seem to be necessary to produce EEs (Bonner 1991, 1995).

EEs can be produced by circularly polarized light (CPL) through asymmetric photochemistry, such as asymmetric photolysis or synthesis (Griesbeck and Meierhenrich 2002; Meierhenrich and Thiemann 2004; Meierhenrich et al. 2005a) as shown in laboratory experiments. Significant EEs (~20%) have been reported in the products of asymmetric photolysis from a racemate (Bonner 1991, 1995). The handedness of EEs via asymmetric photolysis depends on the sign of the CPL due to the difference in photolysis rate of L- and D-enantiomers (Bonner 1991). Even elliptical polarization can induce asymmetric photolysis (Bonner and Bean 2000). The amino acid leucine in the solid state has been photolysed in the laboratory (Meierhenrich et al. 2005b). Furthermore, by irradiation of CPL on interstellar ice analogues, small EEs of less than about 1% have been obtained in laboratory experiments (Nuevo et al. 2006). The possibility of asymmetric synthesis of amino acid precursors in interstellar complex organics using CPL has been demonstrated in a recent experiment by Takano et al. (2007). They prepared complex organic compounds from proton-irradiated gas mixtures as interstellar analogues, and reported EEs of +0.44% by right-circularly polarized UV light and of −0.65% by left-circularly polarized UV light. Amplification of initially low EEs, through autocatalysed reactions, have been experimentally demonstrated (Soai et al. 1995; Shibata et al. 1998; Soai and Kawasaki 2006). Other recent experiments have shown that asymmetric amplification under solid-liquid equilibrium conditions of serine with 1% EE can produce EEs of greater than 99% (Klussmann et al. 2006).

Astronomical sources of CPL that might induce EEs in interstellar material have been investigated. Neutron stars were originally suggested as a possible source of CPL (Rubenstein et al. 1983; Bonner 1991). However neutron stars are not significant sources of CPL at visible and UV wavelengths (Bailey 2001). Bailey et al. (1998) proposed that CPL produced in star-forming regions could contribute to producing the astronomical EEs through asymmetric photolysis. Previous observations indicate that regions of massive star-formation have higher degrees of CPL, although only a relatively small number of star-forming region have been observed (Clayton et al. 2005).

The origin of life and homochirality may be closely related to the formation process for solar-mass stars and their planetary systems. Low mass stars such as the Sun can be formed in massive star-forming regions such as the Orion nebula or relatively isolated regions where only low-mass stars are formed, such as Taurus (Hester and Desch 2005). However, isotopic studies of meteorites that confirm the presence of short half-life radionuclides such as $^{60}$Fe (with a half-life of 1.5 Myr) in the young solar system suggest that a supernova explosion occurred near the Sun (Hester et al. 2004, Hester and Desch 2005, Mostefaoui et al. 2005, Tachibana et al. 2006), indicating the birth of the solar system in a massive star-forming region.

The Orion nebula is the nearest star-forming region in which both high-mass and low-mass stars are being formed (Hillenbrand 1997), and it serves as a valuable test-bed for investigating the CPL mechanism for the origin of EEs. The entire Orion nebula consists of a variety of star forming processes (Genzel and Stutzuki 1989; O'Dell 2001). At the core of the Orion nebula, there is a group of massive (OB-type) young stars, the Trapezium cluster (see also Fig. 1). Around the Trapezium, the Orion nebula harbors the association of many young stars with various mass ranges, the Orion Nebula Cluster (ONC). The embedded massive star-forming region, the BN/KL nebula, is located near the Trapezium. The BN/KL nebula harbors massive protostellar objects such as the BN object and IRc2, with masses of >7 and 25 solar masses, respectively (Genzel and Stutzki 1989). Several young massive stars such as Source I and SMA1 are also thought to exist very close to IRc2 (Gezari 1992; Beuther et al. 2004). The BN object seems to be in an earlier phase of star formation than the Trapezium (Jiang et al. 2005), as well as the deeply embedded sources such as IRc2. The Trapezium stars appear to have evacuated a cavity, near the surface of the molecular cloud OMC-1 (Genzel and Stutzuki 1989; O'Dell 2001). The evacuation of the near-side of the cloud by the Trapezium provides lower extinction to aid observations. Furthermore, background stellar contamination in the Orion nebula is negligible due to the dense molecular cloud behind, and foreground contamination is also relatively low (Jones and Walker 1988; Getman et al. 2005).

As many of the low-mass YSOs will evolve into Sun-like stars, studies of the Orion star-forming region enable us to investigate processes that may have occurred during the birth of our own solar system. In particular, we can explore the circularly polarized radiation that may have bathed the nascent solar system. The obscuring dust prevalent in star-forming regions can be penetrated with observations at near-infrared (NIR) wavelengths which can, thus, be used to study the scattering processes in the circumstellar structures of young stars.

NIR linear polarization (LP) images of the Orion nebula have been reported on a range of scales (e.g., Minchin et al. 1991; Jiang et al. 2005; Simpson et al. 2006). The NIR three color linear polarimetry by Tamura et al. (2006) revealed the extensive (>0.7 pc) LP nebulae around IRc2 and BN. In addition, they reported several small linearly polarized nebulae, the linearly polarized Orion bar, and the low LP near the Trapezium. The LP of hundreds of ONC stars in this region was also investigated, showing the typical hourglass-shaped magnetic field pattern (Kusakabe et al. 2008).

NIR circular polarization (CP) images of the Orion nebula have been reported that reveal a quadrupolar distribution of right- and left-handed CP in the vicinity of OMC-1 (Bailey et al. 1998; Chrysostomou et al. 2000). CP imaging of the Orion BN/KL region show that the quadrupolar structure is centered around the young star IRc2, which appears to be dominant for the large CP (Buschermohle et al. 2005; Fukue et al. 2009). The spatial extent of high CP emission and the degree to which highly polarized radiation interacts with other young stars can only be investigated by extending the spatial coverage of the observations. A first such attempt was reported by Buschermohle et al. (2005), who found generally low degrees of CPL toward several segments of the adjacent HII region. In this paper, we report a deep, wide-field (~6'×6') NIR CP image in the $K_s$ band (2.14 um) of the Orion nebula. Moreover, aperture polarimetry for several hundred point-like sources is also reported. Based on polarimetry results, we discuss possible implications for the origin of EEs, with a view to testing this mechanism for the origin of biological homochirality.

## 2. Observations and Data Reduction

2.14 μm ($K_s$ band) and 1.63 μm ($H$ band) imaging circular polarimetry of M42 were obtained with the SIRIUS camera (Nagayama et al. 2003) and its polarimeter on the 1.4-m IRSF telescope at the South African Astronomical Observatory, on nights during 2006 December. These observations and subsequent data reduction were the same as described in Fukue et al. 2009 (the resultant stellar seeing size ~1.5″), although their observations focus just on the BN/KL region. The estimated uncertainties in the degrees of CPL range from ~0.3% to ~1% close to the corners of the CP image.

2.14 μm ($K_s$ band) imaging linear polarimetry of M42 was obtained with the SIRIUS camera and its polarimeter on the IRSF telescope, on the night of 2005 December 26, with seeing similar to that in the circular polarization observations. These observations and subsequent data reduction were the same as described in Tamura et al. 2006 (see also Kandori et al. 2006; Tamura et al. 2003), with estimated uncertainties less than about 0.3%.

Software aperture circular polarimetry for 540 point-like sources, with intensity signal-to-noise >10, was carried out in a manner similar to that used for linear polarimetry in Kusakabe et al. (2008), and using the same aperture radius of 3 pixels. A total of 353 sources had a polarization signal-to-noise ratio >10 in both the $H$ and $K_s$ bands.

## 3. Results and Discussion of Polarimetry

Fig. 1 shows the wide-field images of circular and linear polarization of the Orion star-forming region in the $K_s$ band (2.14 μm). The field-of-view is 5.5 arcminutes square. The Trapezium is indicated around the center in Fig. 1. The north-west area with strong CP corresponds to the embedded massive star-forming region, the BN/KL nebula, containing the massive protostars IRc2 and BN. These massive protostars are also located near the center of the large LP region in Fig. 1b (Tamura et al. 2006). Point-like sources are not completely cancelled and are visible in the image even if they are unpolarized, because the seeing size changes during the observations of images taken at different quarter-waveplate angles. Since our frame registration is not performed in a sub-pixel unit, the residual stellar profiles on the Stokes $V$ image can be seen as a close pair of positive and negative peaks. This does not affect the polarimetry of extended nebulae on the Stokes $V$ image or the aperture polarimetry of point-like sources performed using each waveplate angle image. The faint circular patterns centered on, and to the south of, the Trapezium in the CP image are ghost images caused by the polarimeter optics.

Our wide-field image in Fig. 1 reveals that the CP region around the BN/KL nebula extends over a large region (up to ~0.4 pc). The degrees of CP are very large, ranging from +17% to −5%, which is consistent with previous polarimetry measurements (Bailey et al. 1998; Chrysostomou et al. 2000; Buschermöhle et al. 2005). The CP map reported in this study covers a much larger area than in previous studies. It reveals that significant CP extends over a region ~400 times the size of the solar system (assumed to be ~200 AU in diameter, including trans-neptunian objects). This extension of the CP region is almost comparable to the size of the linearly polarized region in Fig. 1b (Tamura et al. 2006).

There exists no significant CP around the Trapezium, in contrast with the BN/KL region. In particular, the linearly polarized Orion bar in Fig. 1b (Tamura et al. 2006) shows no significant CP in Fig. 1a. The centrosymmetric LP vector pattern indicates that the polarized Orion bar is irradiated by the Trapezium stars (Tamura et al. 2006). This indicates that the first scattering of the incident radiation from the Trapezium stars by the grains within the bar cannot produce the significant CP; this in turn shows that the dust grains in the LP bar are not well aligned (Gledhill and McCall 2000). The colors of this region show that the Trapezium and the bar are located near the surface of the cloud (Buschermöhle et al. 2005) in contrast with the BN/KL region.

Most of the low- or medium-mass young stars in Fig. 1 do not show extended structure in either LP or CP, in contrast to the BN/KL region. Even those with a NIR nebula that is linearly polarized (e.g., OMC-1S, see Tamura et al. 2006; see also Fig. 1), show no significant CP, even when the nebula is spatially resolved. Fig. 2 shows the distribution of the aperture circular polarimetry, for the 353 point-like sources detected both in the $K_s$ band and $H$ band with a polarization signal-to-noise ratio >10. Many of these sources are low-mass young stars whose circumstellar structures are unresolved at a 1.5-arcsecond resolution (equivalent to about 700 AU). Fig. 3 shows a *J-H* vs. *J* color-magnitude diagram for these sources. In the diagram, we indicate the locus of 1 Myr old pre-main sequence (PMS) stars at the distance of 460 pc based on the stellar evolution model by Testi et al. (1998). Also we indicate the reddening direction based on Cohen et al. (1981). The diagram is consistent in indicating that these sources are 1-Myr old PMS stars with masses less than ~3 solar masses. The vast majority of these sources measured in this study are cluster members (Jones and Walker 1988; Getman et al. 2005; Hillenbrand 1997; Lucas et al. 2001). The proper motions and radial velocities of ONC members show a dispersion of a few km s$^{-1}$ (Jones and Walker 1988; Fűrész et al. 2008), implying that these stars will move within about 1 pc, in 1 Myr. In Fig. 2, the measured degree of CP for each source is generally small. We conclude that none of the detected point sources clearly show significant integrated circular polarizations (>than 1.5 % both in $K_s$ and $H$ bands in the same handedness); one source does have a CP > 1.5%, both in the $K_s$ and $H$ bands, but is embedded in the western high CP region and hence substantially contaminated. OMC-1S shows aperture circular polarimetry of about 0.3% in $K_s$ band. These results are consistent with previous observations (Clayton et al. 2005).

## 4. CP in Massive Star-forming Regions: Possible Implications for the Origins of Homochirality

We will now discuss the implications of these results for the origin of biomolecular homochirality. Bailey (2001) discusses how CPL in star-forming regions might be important in producing EEs and ultimately seeding homochirality on terrestrial planets. Imaging circular polarimetry of several YSOs (Gledhill et al. 1996; Chrysostomou et al. 1997; Bailey et al. 1998; Chrysostomou et al. 2000; Clark et al. 2000; Ménard, et al. 2000; Chrysostomou et al. 2007; Fukue et al. 2009; Clayton et al. 2005) and numerical simulations (Fischer et al. 1996; Wolf et al. 2002; Whitney and Wolff 2002; Lucas et al. 2004; Lucas et al. 2005; Chrysostomou et al. 2007) indicate that a YSO will not usually generate a net CP because it will have regions of positive and negative sign that cancel globally. Hence, a nascent solar system around a low-mass star would not be irradiated by a net CP. A low-mass YSO would only experience strong CP of a single sign when it is externally irradiated by a high-mass YSO. In our polarimetry results, low-mass young stars themselves do not show strong one-handed CP. On the other hand, extended regions of high CP (hundreds of times the size of the solar system) are associated with high-mass stars. Large numbers of low-mass YSOs are often located in a clustered star-forming region containing massive stars. The high stellar density (>10$^3$ stars pc$^{-3}$) and the large and wide CP region around the location of IRc2 suggest that there are at least several stars in the high CP region around IRc2. There, a low-mass young star can see predominantly one-handedness of CP, which provides an external source for asymmetric photolysis to yield EEs in any chiral molecules (Bailey 2001, Bonner 1991).

Photolysis of amino acids requires UV radiation, rather than the infrared radiation observed in this study. UV radiation cannot be directly observed as it is unable to penetrate the dust that lies along the line-of-sight between the Earth and regions of high CP. Numerical calculations (Bailey et al. 1998) indicate that significant amounts of UV CP can be produced by young stars and this could spread over large distances because of the large cavities formed by bipolar outflows and jets (Tamura et al. 2006). UV CP can then be produced by mechanisms discussed by Lucas et al. (2005).

Should the asymmetric photochemical processes reported in laboratory experiments operate in regions of high-mass star-formation, then they could give rise to the observed EEs of meteoritic amino acids, possibly amplified through autocatalysis. Assuming that the observed EEs were produced in the nascent solar system, the detection of EEs of meteoritic amino acids on Earth suggests that the EEs can survive for many billions of years. Our observation of wide regions of high CP suggests that similar CP could have irradiated the early solar system if it formed in a similar environment. Recently, Glavin and Dworkin (2009) have detected no L-isovaline excess for the most pristine Antarctic CR2 meteorites Elephant Moraine 92042 and Queen Alexandra Range 99177, whereas they have detected large L-EEs in the CM meteorite Murchison and the CI meteorite Orgueil. They discuss the possibility that the detected EEs may be produced by amplification of small initial EEs during an aqueous alteration phase.

The high spatial extent of large degrees of CPL, together with the various laboratory experiments, supports the idea that the initial seeds of homochirality are generated in the nascent solar system and are carried to Earth during the heavy bombardment that occurred in the Earth's early history (Bailey et al. 1998), with subsequent chiral amplification (Barron 2008, Soai and Kawasaki 2006, Klussmann et al. 2006).

The above scenario for the origin of homochirality requires that the Sun be formed near a cluster of massive stars (Hester and Desch 2005), which produce most of the scattered light over a very large region of a few pc. Stars can also form in relative isolation in a molecular cloud that forms only low-mass stars. That the solar system originated in a massive star formation region is supported by isotopic studies of meteorites such as $^{60}$Fe suggesting that a supernova explosion occurred near the Sun (Mostefaoui et al. 2005, Tachibana et al. 2006). The possibility that the solar protoplanetary disk survived even a supernova explosion is supported by numerical simulations (Ouellette et al. 2007).

## 5. Conclusion

CPL, produced in regions of high-mass star-formation, is one possibility for producing EEs in small bodies in the presolar nebula, which could then be delivered to the early Earth, thereby contributing to the evolution of homochirality in living organisms. NIR wide-field (~6'×6') imaging circular polarimetry of the core of the Orion nebula show that high CP extends to ~0.4 pc around the massive star-forming region, the BN/KL nebula. This extension of CP is comparable with that of LP. On the other hand, the area other than the massive star forming region generally showed low CP, and most of the low- or medium-mass young stars do not show detectable extended structure associated with them in either LP or CP, in contrast to the BN/KL region. Even OMC-1S, having a NIR nebula indicated by the extended circumstellar structures in the LP map, shows no extensive regions with significant CP, and has very low CP measured through aperture polarimetry. The aperture polarimetry of several hundred point-like sources showed low CP, indicating that low- or medium-mass young stars (i.e., sun-like stars) themselves do not show significant CP. If our solar system formed in a massive star-forming region (not in a low mass star-forming region) and was irradiated by asymmetric CP, then EEs could have been produced in the parent bodies of the meteorites delivering an initial chiral bias of amino acids (or precursor) onto the early Earth.

**Acknowledgements** We acknowledge discussions with T. Nagata, T. Nagayama, and S. Sato. We thank F. Palla for providing us with the table of the stellar model of Testi, Palla & Natta (1998). T.F. was supported by Research Fellowships of the Japan Society for the Promotion of Science (JSPS) for Young Scientists. This work was partially supported by KAKENHI 18-3219. M.T. is supported by Grants-in-Aid from the Ministry of Education, Culture, Sports, Science and Technology (MEXT) of Japan (16077101, 16077204), and that from the JSPS (19204018). D.C.B.W. acknowledges support from the NASA Exobiology Program (grant NNX07AK38G) and the NASA Astrobiology Institute. IRAF is distributed by the National Optical Astronomy Observatories, which are operated by the Association of Universities for Research in Astronomy, Inc., under cooperative agreement with the National Science Foundation.

**Fig. 1** Image of degree of polarization (%) in the $K_s$ band (2.14 μm) of the central region of the Orion star-forming region. **a** Image of circular polarization degree; **b** The degree of linear polarization. The field-of-view is 5.5 arcminutes or 0.74 pc square at a distance of 460 pc. North is up and east is to the left. The positions of IRc2 and BN are indicated by a cross and a circle, respectively, while those of the Trapezium stars and the low-mass young star OMC-1S are denoted by big and small arrows, respectively. A positive sign for CP indicates that the electric vector is rotated anticlockwise in a fixed plane relative to the observer

**Fig. 2** Histograms of circular polarization degree (%) of 353 point-like sources. **a** in the $K_s$ band (2.14 μm); **b** in the $H$ band (1.63 μm). The histograms are constructed using a bin width of 0.2%

**Fig. 3** Color-magnitude diagram for 353 point-like sources used in Fig. 2, using their *J*-band (1.25 μm) and *H*-band (1.63 μm) data in the same observation. The vertical axis shows *J* magnitude, and the horizontal axis shows *J-H* magnitude. Our observational data are plotted with crosses. The filled circles denote the loci of 1 Myr old PMS stars at 460 pc, according to the stellar evolution model by Testi et al. (1998). The assumed masses are 0.1, 0.2, 0.4, 0.6, 0.8, 1, 1.2, 1.5, 2, 2.5, 3, and 3.5 solar masses, from bottom to top (the second point from the top for 3.5 solar masses), connected by the solid line. The dashed line identifies the reddening law through the loci of the 2.5 solar masses (Cohen et al. 1981)

**Fig.1**

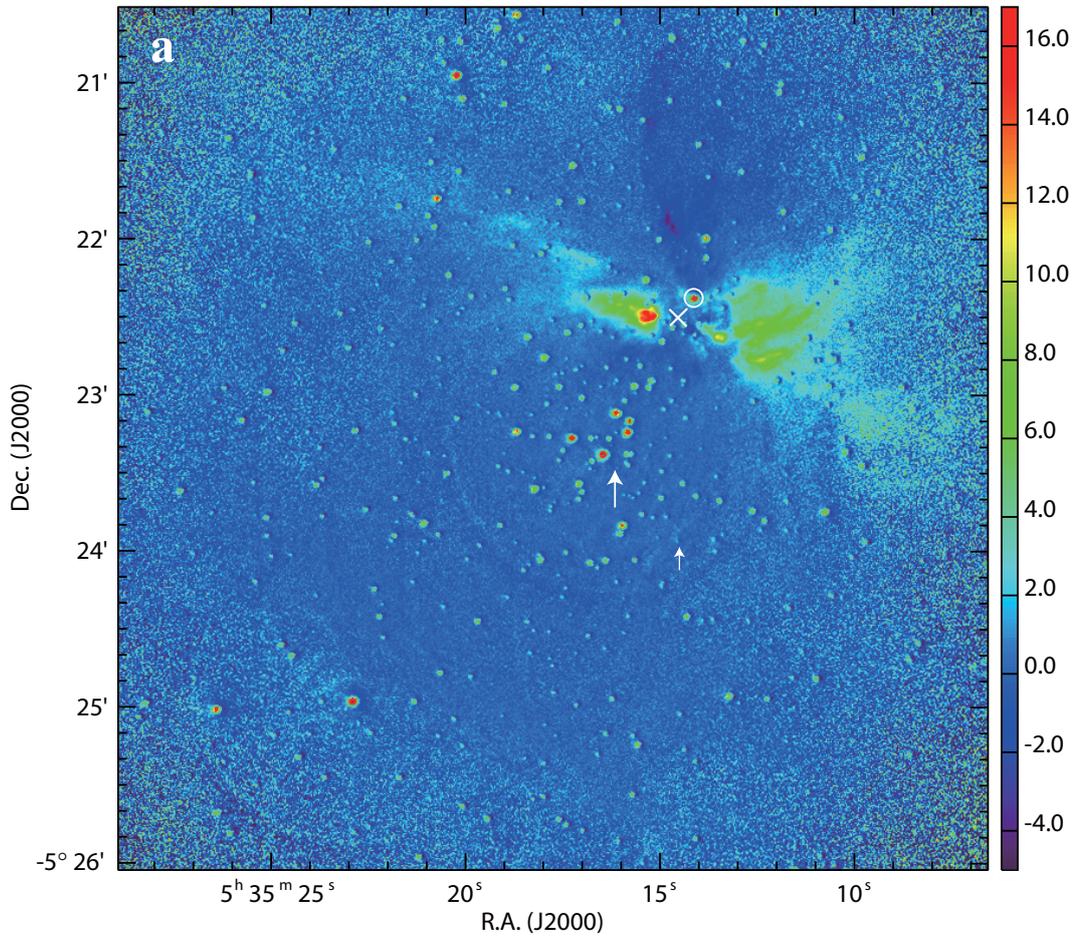

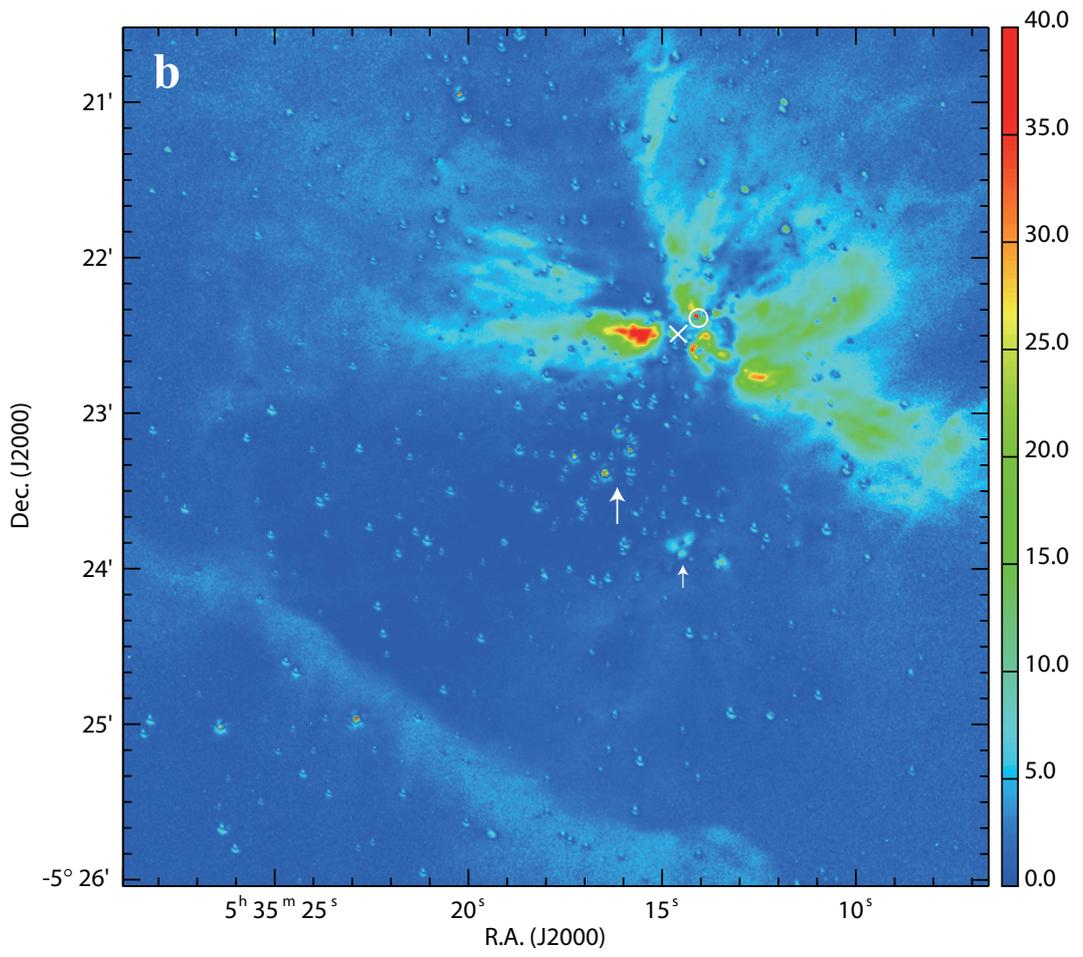

**Fig.2**

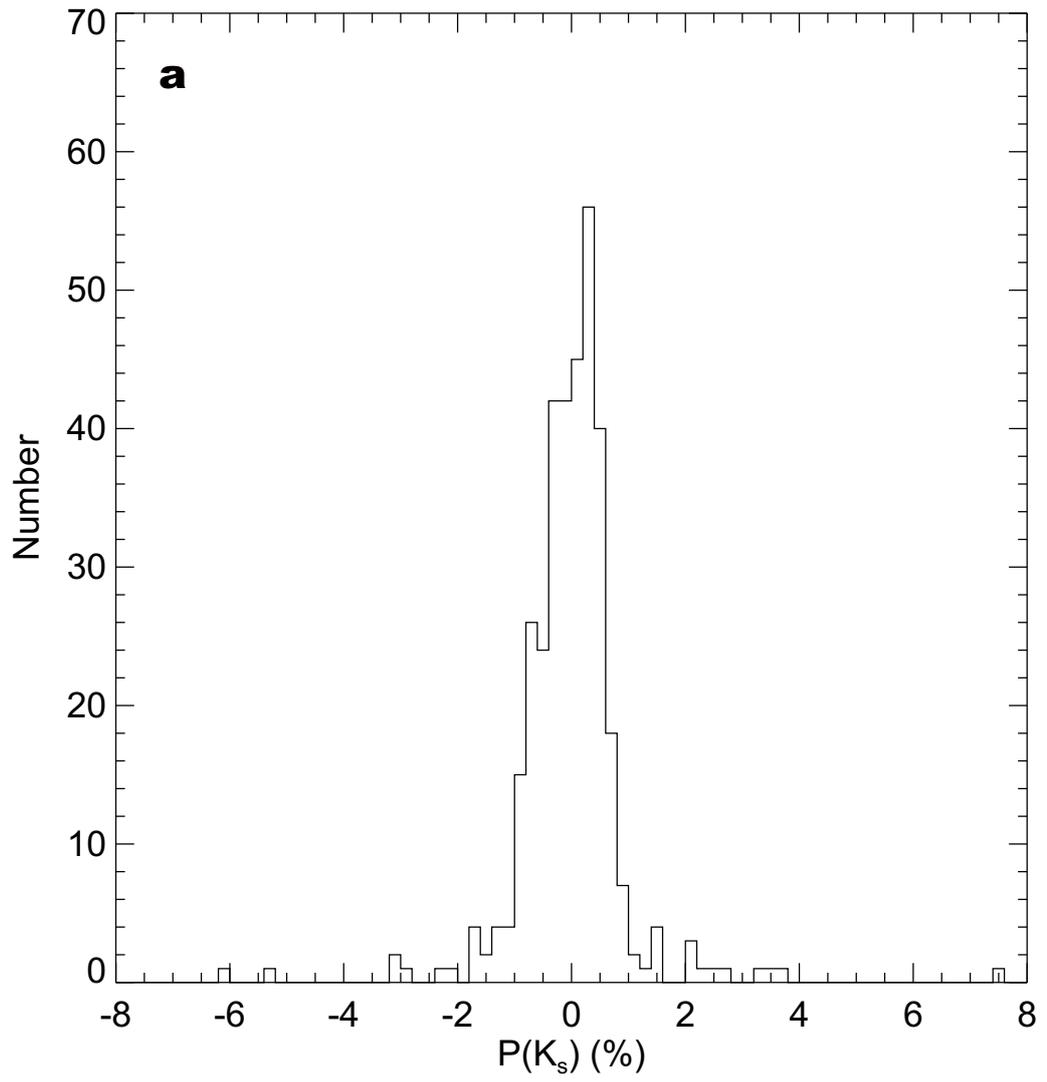

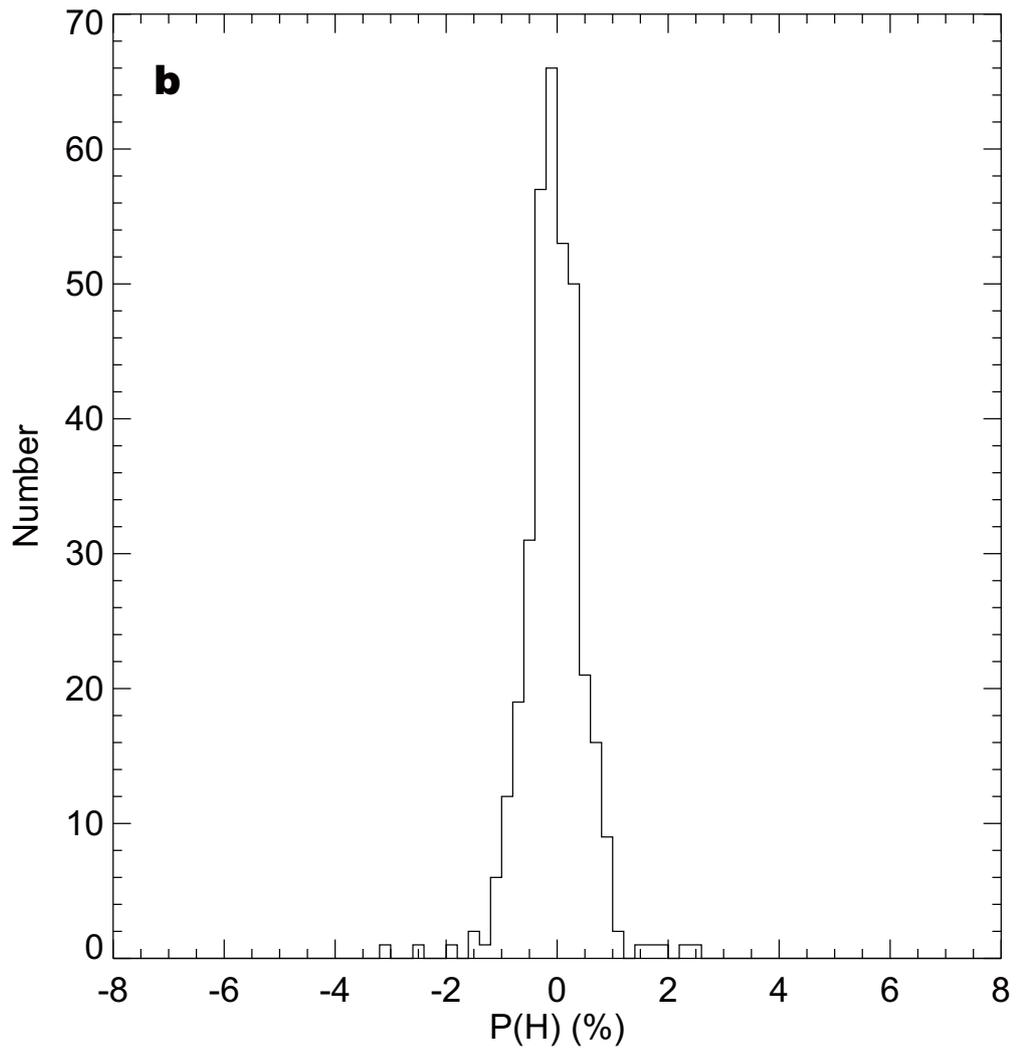

**Fig. 3**

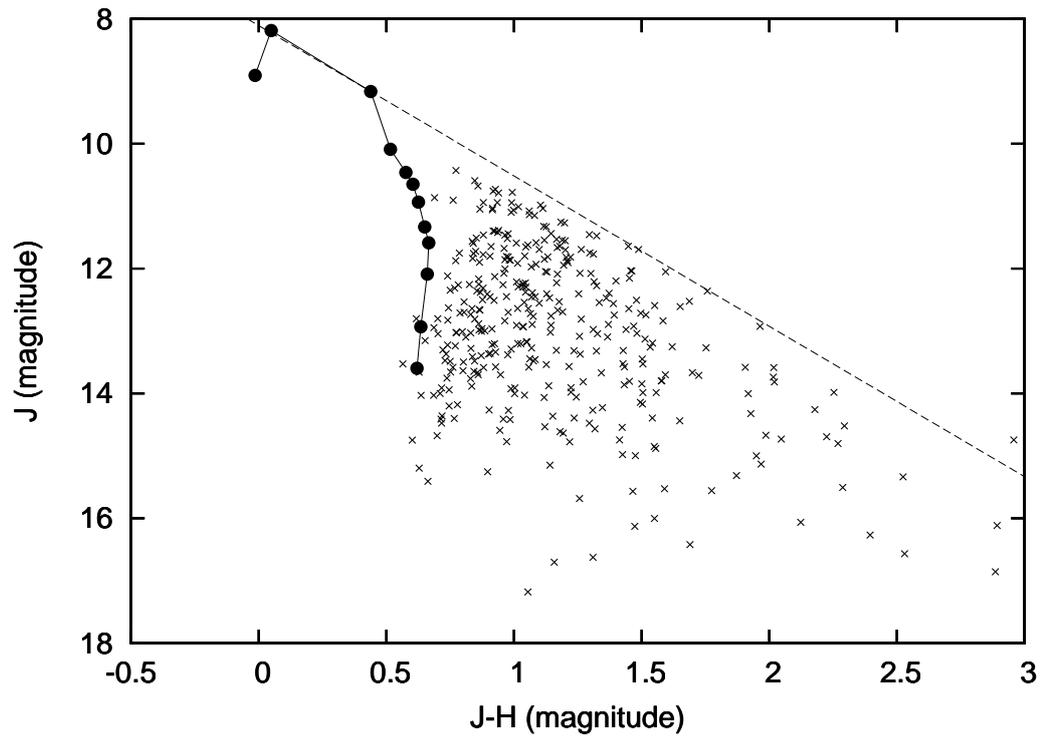